\documentstyle[pra,aps,twocolumn,epsfig]{revtex}
\newcommand{\rr}{{\bf r}}
\newcommand{\rrp}{{\bf r'}}
\newcommand{\kk}{{\bf k}}
\newcommand{\nn}{{\bf n}}
\newcommand{\nnp}{{\bf n'}}
\begin{document}

\wideabs{

\title{Ground state and elementary excitations of single and 
binary \\ Bose-Einstein condensates of trapped  dipolar gases}
\author{K. G\'oral$^{1,2}$ and L. Santos$^{1}$}
\address{
(1) Institut f\"ur Theoretische Physik, Universit\"at Hannover, D-30167 
Hannover, Germany\\ 
(2) Center for Theoretical Physics, Polish Academy of Sciences, \\ Aleja 
Lotnik\'ow 32/46, 02-668 Warsaw, Poland 
}
\maketitle

\begin{abstract}
We analyze the ground-state properties and the excitation spectrum of 
Bose-Einstein condensates of trapped dipolar particles. 
First, we consider the case of 
a single-component polarized dipolar gas. For this case we discuss the
influence of the 
trapping geometry on the stability of the condensate as well as the 
effects of the dipole-dipole interaction on the excitation spectrum. 
We discuss also the ground state and excitations of 
a gas composed of two antiparallel dipolar components.
\end{abstract}

}

\section{Introduction}
\label{sec:introd}

The nature and stability of a Bose-Einstein condensate (BEC) \cite{BEC,Li} 
are strongly influenced by the interparticle interactions, which are 
described  by the $s$-wave scattering length $a$. 
If $a>0$ the interactions are repulsive, and condensates with 
an arbitrary large number of particles are stable. 
On the contrary, spatially homogeneous condensates with $a<0$ 
are absolutely unstable with regard to local collapses \cite{Dalfovo}. 
The presence of a trapping potential changes the situation drastically, as 
revealed in successful experiments with magnetically 
trapped atomic $^7$Li ($a=-14$ \AA) \cite{Li,collapse}.
As found in theoretical studies \cite{keith,baym,negat}, 
there will be a metastable BEC if the 
number of condensed particles is sufficiently small, 
such that the spacing between the trap levels exceeds 
the mean field interparticle interaction $n_0|g|$ (where $n_0$ is the
condensate density, $g=4\pi\hbar^2a/m$, 
and $m$ is the atom mass). In other words, the BEC is stabilized 
if the negative pressure
caused by the interparticle attraction is compensated by the quantum 
pressure imposed by the trapping potential.

The effects of the interparticle interactions on the condensate 
properties have been mainly discussed for the case of van der Waals 
(short-range) interactions.
However, the BEC in the presence of dipole-dipole 
interactions has recently raised a considerable interest 
\cite{Yi1,Goral,Santos,Yi2,recyou,Martikainen,Meystre,Giovanazzi,kurizki}. 
Novel physics is 
expected for dipolar 
BEC, since the dipole-dipole interactions are long-range, anisotropic and 
partially 
attractive. The non-trivial task of achieving and controlling dipolar BECs 
is thus particularly challenging.

The interest on dipolar gases has been 
partially motivated by the recent success in creating ultra-cold
molecular clouds \cite{molecules,Heinzen}. This success  
opens fascinating prospects to achieve 
quantum degeneracy in trapped gases of heteronuclear molecules, which  
could interact via electric dipole-dipole forces after being 
oriented in a sufficiently high electric field.
On the other hand, the ultra-cold gases of 
atoms with large magnetic dipole moments, as chromium 
\cite{Weinstein,Pfau1,Celotta,Pfau2,EvapCr} 
or europium \cite{europium}, have also been subject of growing 
interest. In this case, the dipole-dipole interactions are not expected to 
be dominant, 
although for a relatively small $a$ the BEC may reflect the interplay 
between short-range and dipole-dipole interactions 
\cite{Goral,Martikainen}. Interestingly, these effects can 
be amplified by reducing $a$ via Feshbach resonances 
\cite{FeshbachMIT,85Rb}.  
Similar effects have been discussed in Refs.\ \cite{Yi1,Yi2,recyou} for 
ground-state atoms with  
electric dipole moments induced by a high dc electric fields (of the order
of 10$^6$ V/cm). It has been also suggested that laser-induced 
dipole-dipole interactions could be achieved by exciting atoms to Rydberg 
states \cite{Santos}. In this case applications to quantum information 
processing have been discussed (see e.g. \cite{Lukin}).

The stability of the condensate is significantly modified 
by the presence of dipole-dipole interactions 
\cite{Yi1,Goral,Santos,Yi2,Martikainen}. 
In particular, a BEC of particles dominantly interacting via dipole forces 
is, 
similarly to condensates with $a<0$,
unstable in a spatially homogeneous case and can be stabilized by
confinement in a trap. It has been shown \cite{Santos} that 
the sign and the value of the 
dipole-dipole interaction energy is strongly influenced 
by the trapping geometry and, hence, the stability diagram depends 
crucially 
on the trap anisotropy. This opens  new interesting possibilities for 
controlling and 
engineering macroscopic quantum states. In particular, for dipoles 
oriented 
along the axis of a cylindrical trap there exists a critical
value $l_*=0.4$ for the square root of the 
ratio of the radial to axial frequency
$l=(\omega_{\rho}/\omega_z)^{1/2}$. 
Pancake traps with $l<l_*$  provide mostly  
a repulsive mean field of the dipole-dipole interaction, and thus the
dipolar condensate in such traps will be stable at any number of particles 
$N$. For $l>l_*$ the stability requires $N<N_c$, where
the critical value $N_c$ at 
which the collapse occurs is determined by the condition that (on
average) the mean field interaction energy is attractive and  close in 
absolute value to 
$\hbar\omega_{\rho}$.

The study of the condensate properties would be incomplete without 
the analysis of the excitation spectrum, which determines 
the dynamical behavior of the system in the regime of weak perturbations.
Zero-temperature excitation frequencies have been extensively studied in
the case of condensates in dilute alkali gases, both experimentally
\cite{JILAexc,MITexc} and theoretically \cite{exc_th,Perez,tom}. 
In this paper we discuss the low-lying
collective excitation frequencies of trapped dipolar BEC, 
complementing the analysis of Refs.\ \cite{Yi2,recyou}.  
We first consider the case of dominant dipole-dipole interactions, and 
later on 
we discuss the situation where the short-range interaction is also 
relevant. 
In particular, we discuss in detail the nature of instability 
and demonstrate that one of the excitations frequencies tends to zero at 
the 
criticality as a power of $(N_c-N)^{\beta}$. 
We discuss the qualitative character of the low-lying modes, and show that 
the exponent $\beta$ undergoes 
a crossover from $1/4$ for $l\gg l_*$ to $2$ at $l\simeq l_*$.
To illustrate the case of mixed short-range and dipole-dipole 
interactions, we present 
predictions for the excitation frequencies for the particular case of a 
chromium gas.

The second part of the paper is devoted to the analysis of binary dipolar 
BECs.
In recent years the development of trapping techniques has allowed for 
creation of multicomponent condensates, formed by 
atoms in different internal (electronic) states
\cite{binaryJILA,binaryMIT}. The physics of multicomponent BEC,   
far from being a trivial extension of the single-component one, 
presents novel and fundamentally different scenarios for its ground-state  
wave function \cite{binary-ground,binaryTrippenbach} and excitations
\cite{binary-exc}. 
In particular, it has been experimentally observed that a BEC can reach   
an equilibrium state characterized by phase separation of the species in  
different domains \cite{binaryMIT}.
The analysis of multicomponent BEC has been so far mostly limited to the
case  of short-range interparticle interactions (a model long-range
interaction has been considered in Ref. \cite{binaryTrippenbach}). One of
the main goals of this paper is to analyze the properties of 
multicomponent 
dipolar BEC. Such a mixture can be achieved in
different physical contexts. In particular, it would be in general the
case for experiments in ultracold polar molecules \cite{Meijerpriv}, and 
in chromium \cite{Pfaupriv} in which different magnetic-moment 
species are simultaneously trapped in an optical dipole trap. 
It would also be the case of atomic electric dipoles created by
laser-induced pumping to two different Rydberg states. 
Finally, the same situation would appear in condensates of heteronuclear
Hund A diatomic molecules, for which the direction of the magnetic moment 
is
correlated (parallel or antiparallel) with the direction of the
molecular axis. Thus, if the magnetic moments are oriented in a magnetic 
field, the  
electric moments can acquire two possible directions. 
We show below that a binary dipolar BEC of two antiparallel dipole 
components 
differs qualitatively from the case of a short-range interacting binary 
BEC.

Our paper is organized as follows. In Sec.\ \ref{sec:gr1} we briefly 
review 
the ground-state properties of a single component BEC of trapped dipolar
gases \cite{Santos}.  Sec.\ \ref{sec:exc1} is devoted to the analysis 
of the  excitation spectrum of single component trapped dipolar condensed 
gases. Sec.\ \ref{sec:exp} briefly discusses the ballistic expansion of a 
dipolar BEC. 
In Sec.\ \ref{sec:gr2} the ground state  of a BEC of two different dipolar 
species
is considered. The excitation  spectrum for this case is discussed in
Sec.\ \ref{sec:exc2}. We conclude in Sec.\ \ref{sec:concl}.

\section{Ground-state properties of a single-component dipolar BEC} 
\label{sec:gr1}

\subsection{Description of the system}

In this section we briefly review the results of Ref.\ \cite{Santos}. 
We consider a condensate of dipolar particles in a cylindrical harmonic trap. 
In the following we consider the case of electric dipoles, although the same 
physics is expected for magnetic ones \cite{Goral}.
All dipoles are assumed to be oriented along the trap axis by a sufficiently 
strong external field. Accordingly,  
the dipole-dipole interaction potential between two dipoles is given by 
$V_d({\bf R})=(d^2/R^3)(1-3\cos^2{\theta})$, 
where $d$ characterizes the dipole moment,
${\bf R}$ the vector between the dipoles ($R=|{\bf R}|$ being its length),
and $\theta$ the angle between ${\bf R}$ and the dipole orientation. 
Similarly as in Refs.\ \cite{Yi1,Goral,Santos},
we describe the dynamics of the condensate wave function $\psi({\rr},t)$
by using the time-dependent Gross-Pitaevskii equation (GPE):
\begin{eqnarray}
&& i\hbar\frac{\partial}{\partial t}\psi({\rr},t)=  
\left \{
-\frac{\hbar^2}{2m}\nabla^2+\frac{m}{2}(\omega_{\rho}^2\rho^2+\omega_z^2z^2)+
\right \delimiter 0 \nonumber \\
&& \left \delimiter 0 + g|\psi({\rr},t)|^2 
+d^2\int d{\rrp} 
\frac{1-3\cos^2\theta}
{|{\rr}-{\rrp}|^3}
|\psi({\rrp},t)|^2 \right \}
\psi({\rr},t). 
\label{GPE1}
\end{eqnarray}
where $\psi({\rr},t)$ is normalized to the total number of condensate 
particles $N$. The third term in the rhs of Eq.\ (\ref{GPE1}) 
corresponds to the mean field of short-range (van der Waals) forces and 
the last 
term to the mean field of the dipole-dipole interaction. 
In this section, we omit the term $g|\psi({\rr},t)|^2\psi({\rr},t)$,
assuming that the interparticle interaction is dominated by dipole-dipole
forces ($d^2\gg |g|=4\pi\hbar^2|a|/m$) and the system is away from shape
resonances of $V_d({\bf R})$. The effects of the short-range interactions 
on the excitation spectrum are discussed in detail in Sec.\ \ref{sec:exc1}.

The wave function of the relative motion of a pair of dipoles is influenced by the 
dipole-dipole interaction at interparticle distances $|{\rr}-{\rrp}|\alt
r_*=2md^2/\hbar^2$. This influence is ignored in the dipole-dipole term of
Eq.(\ref{GPE1}), as the main contribution to the integral comes from distances
$|{\rr}-{\rrp}|$ of order the spatial size of the
condensate, which we assume to be much larger than $r_*$.

Away from the shape resonances the dipolar condensate is unstable in the 
spatially homogeneous case. For all dipoles
parallel to each other, by using the Bogoliubov method, one finds 
an anisotropic dispersion law for elementary excitations:
$\varepsilon({\bf k})=[E_k^2+8\pi E_kn_0d^2(\cos^2{\theta_k}-1/3)]^{1/2}$,
where $E_k=\hbar^2k^2/2m$, $n_0$ is the condensate density, and $\theta_k$
is the angle between the excitation momentum ${\bf k}$ and the direction
of
the dipoles. The instability is clearly seen from the fact that at small
$k$ and $\cos^2{\theta_k}<1/3$ imaginary excitation energies 
$\varepsilon$ emerge.  

\subsection{Numerical results}

Equation (\ref{GPE1}), contrary to the usually employed GPE with 
short-range interactions, is an integro-differential equation. The 
evaluation of the integral term deserves special attention, since the 
integrand diverges at relative interparticle distances tending to zero. 
Fortunately, the calculation of the integral term can be simplified by 
means of the convolution theorem:
\begin{equation}
d^2\int d{\rrp} 
\frac{1-3\cos^2\theta}
{|{\rr}-{\rrp}|^3}
|\psi({\rrp})|^2 =
{\cal F}^{-1}\left\{ {\cal F}[V] ({\bf q}){\cal F}[|\psi|^2]({\bf q})
 \right\},
\label{convo}
\end{equation}
where ${\cal F}$ and ${\cal F}^{-1}$ indicate the Fourier transform and 
the inverse Fourier transform, respectively. The Fourier transform of the 
dipole-dipole potential reads:
\begin{equation}
{\cal F}[V]({\bf q})=4\pi d^2 (1- 3\cos^{2} \alpha)\; [
\frac{\cos(bq)}{(bq)^{2}} - \frac{\sin(bq)}{(bq)^{3}} ],
\end{equation}
where $\alpha$ is the angle between the momentum $\bf q$ and the dipole 
direction, and $b$ is a cutoff distance corresponding to the atomic 
radius (a few Bohr radii). Since $b$ is much smaller than any significant 
length scale of the system, one can safely perform the limit 
\begin{equation}
\lim_{b \rightarrow 0} {\cal F}(V({\bf r})) = \frac{4\pi}{3}
d^{2}\;(3\cos^{2} \alpha-1).
\end{equation}
In order to evaluate the Fourier transform of Eq. (\ref{convo}) ${\cal F}(|
\psi|^{2})$ is numerically evaluated by means of a standard Fast Fourier 
Transform (FFT) algorithm and multiplied by ${\cal F}(V({\bf r}))$.

The ground-state of the system is obtained by employing a standard 
split-operator technique in imaginary time. The split-operator technique 
is also based in an FFT algorithm and, consequently, for each time step 
four FFT's are needed: two for the calculation of the integral term and 
two for the evolution. We would like to stress that this procedure 
constitutes a non trivial computational task. Additionally, the FFT 
algorithm must be evaluated in Cartesian coordinates and, as a 
consequence, computationally demanding fully three-dimensional 
calculations are required.

To understand the influence of the trapping potential on the
dipolar condensate, we have simulated Eq.(\ref{GPE1})
for various values of the number of particles $N$, dipole moment $d$, and
the trap aspect ratio $l$. We have found the conditions under which the condensate 
is stabilized by the trapping field and investigated static properties of this
Bose-condensed state. 

For a stationary condensate the wave function 
$\psi({\rr},t)=\psi_0({\rr})
\exp{(-i\mu t/\hbar)}$, where $\mu$ is the chemical potential, and the lhs of
Eq.\ (\ref{GPE1}) becomes $\mu\psi_0({\rr})$. The important energy scales
of
the problem are the trap frequencies $\omega_z$, $\omega_{\rho}$ and the
dipole-dipole interaction energy per particle defined as
$V=(1/N)\int V_d({\rr}-{\rrp})\psi_0^2({\rr})\psi_0^2({\rrp})d{\rr}
d{\rrp}$.
Accordingly, the trap frequencies, 
and the (renormalized) number of particles  
$\sigma=Nr_*/a_{\rm max}$ (with $a_{\rm max}=
(\hbar/2m\omega_{\rm min})^{1/2}$ 
being the maximal oscillator length of the trap) 
form the necessary set of parameters 
allowing us to determine the chemical potential and give a full description
of the ground state of a trapped dipolar condensate.

We have found that the dipolar condensate is stable
either at $V>0$ or at $V<0$ with $|V|<\hbar\omega_{\rho}$. This
requires $N<N_c$, where the critical number $N_c$
depends on the trap aspect ratio $l=(\omega_{\rho}/\omega_z)^{1/2}$.
The calculated dependence $N_c(l)$ 
clearly indicates the presence of a critical point $l_*\simeq 0.43$\cite{Santos} . 
In pancake traps with $l<l_*$ the condensate is stable at any $N$, because
$V$ always remains positive. 
For small $N$ the shape of 
the cloud is Gaussian in all directions. With increasing $N$,
the quantity $V$ increases and the cloud first becomes Thomas-Fermi in the
radial direction and then, for a very large $N$, also axially. The  
ratio of the axial to radial size of the cloud, $L=L_z/L_{\rho}$, continuously 
decreases with increasing
number of particles and reaches a limiting value at $N\rightarrow\infty$
(see Fig. 3 of Ref.\cite{Santos}).  In this respect, for a
very large $N$ 
we have a pancake Thomas-Fermi condensate. 

For $l\geq 1$ the mean field dipole-dipole interaction is always 
attractive.
The quantity $|V|$ increases with $N$ and the shape of the cloud changes. 
In spherical traps the cloud becomes more elongated
in the axial direction and near $N=N_c$ the shape
of the cloud is close to Gaussian with the aspect ratio 
$L=2.1$. In cigar-shaped traps ($l\gg 1$) 
especially interesting is the regime
where $\hbar\omega_z\ll |V|\ll\hbar\omega_{\rho}$. In this case the radial
shape of the cloud remains the same Gaussian as in a non-interacting gas,
but the axial behavior of the condensate will be governed by the 
dipole-dipole interaction which acquires a quasi 1-dimensional (1D) 
character. Thus, one has
a (quasi) 1D gas with attractive interparticle interactions and
is dealing with a stable (bright) soliton-like condensate where attractive
forces are compensated by the kinetic energy \cite{soliton}. 
With increasing $N$, $L_z$ decreases. Near $N=N_c$, where
$|V|$ is close to $\hbar\omega_{\rho}$, the axial shape of the cloud also
becomes Gaussian and the aspect ratio takes the value $L\approx 3.0$.     
For $l_*\leq\l<1$ the dipole-dipole interaction energy is positive for a 
small number of particles and increases with $N$. The quantity $V$
reaches its maximum and further increase in $N$ reduces $V$ and makes
the cloud less pancake-shaped. At the critical point $N=N_c$ the shape of 
the cloud is close to Gaussian and the aspect ratio $L<3.0$.

In the previous analysis, the case of dominant dipole-dipole interactions 
was considered. However \cite{Goral,Martikainen,Yi1,Yi2}, in the general 
case the effects of the short-range interactions are comparable 
or even larger than those related to the dipole-dipole interactions. 
In  such situations the short-range term must be maintained in Eq.\ (\ref{GPE1}). 
Provided that $a$ is sufficiently small and positive, the system
can become unstable and undergo a collapse in a similar way to what was
observed in experiments with $^{7}$Li \cite{Li} and $^{85}$Rb
\cite{85Rb}. For the case of negative $a$ the dipolar gas is expected to be highly unstable, but 
the dipole-dipole interaction could be employed to stabilize the gas by the
trap geometry in a way analogous to the one presented in Ref. \cite{Santos}.

\section{Excitations of a single-component dipolar BEC}
\label{sec:exc1}

In this section we analyze the collective excitations of a 
dipolar BEC. Since this is a potentially unstable system, there
is a fundamental question about the qualitative and quantitative 
nature of this instability. Another important question concerns 
quantitative aspects, in particular how relevant are the effects of the 
dipole-dipole interaction in the excitation spectrum and to what extent 
they can observed in the experiments.

In the subsection \ref{sec:exc1a}, we 
briefly summarize the variational approach 
introduced by P\'erez-Garc\'\i a {\it et al.} \cite{Perez} and later used 
by Yi and You \cite{Yi2,recyou} to describe the low-lying excitation
spectrum  of a dipolar condensate (for a different variational method 
using the self-similarity assumption see \cite{self-similar}).
We employ the notation introduced in the Refs.\ \cite{Yi2,recyou}.
Subsection \ref{sec:exc1b} is devoted to the analysis of the behavior of the excitations close to the 
instability. In subsection \ref{sec:exc1c}, 
we calculate numerically the response of the system 
to small perturbations and compare the results with those yielded by the 
method of 
subsection \ref{sec:exc1b}. 
Finally, in subsection \ref{sec:exc1c}, we also discuss 
how the effects of the dipole-dipole interactions can manifest themselves 
in the excitation spectrum for various kinds of dipolar gases and trapping 
geometries.

\subsection{Variational approach}
\label{sec:exc1a}

The problem of solving the time dependent GPE Eq. (\ref{GPE1}) can be 
restated as a
variational problem corresponding to the stationary point of the action  
$S=\int dt d{\rr} {\cal L}$ related to the Lagrangian density
\cite{Yi2,recyou,Perez}:
\begin{eqnarray}
{\cal L}&=&{i\over
2}\hbar\left[\psi({\rr}){\partial\psi^*({\rr})\over\partial
t}-\psi^*({\rr}){\partial\psi({\rr})\over\partial t}\right]\nonumber\\
&+&{\hbar^2\over 2m}|\nabla\psi({\rr})|^2+V_{\rm
t}({\rr})|\psi({\rr})|^2\nonumber\\
&+&{g\over 2}|\psi({\rr})|^4+{u_2\over 2}|\psi({\rr})|^2\int
d{\rrp}{Y_{20}(\theta)\over |{\rr}-{\rrp}|^3}|\psi({\rrp})|^2 \; ,
\label{Ldens}
\end{eqnarray}
where $u_2=4\sqrt{\pi/5}d^2$ and $V_{\rm t}({\rr})$ describes the
trapping potential. 
We consider the following Gaussian ansatz for the condensate wave
function: 
\begin{eqnarray}
\psi(x,y,z,t)=A(t)\prod_{\eta=x,y,z}e^{-\eta^2/2w^2_\eta}e^{i\eta^2\beta_\eta(t)},
\label{Ansatz}
\end{eqnarray}
where in contrast to Ref.\ \cite{Yi2} we omit the
possibility of the sloshing motion of the condensate. By inserting the
ansatz (\ref{Ansatz}) into Eq. (\ref{Ldens}) and integrating over the 
spatial
coordinates, one obtains the effective Lagrangian. 
From this effective Lagrangian one finds the equations of
motion of the variational parameters, i.e. the corresponding
Euler-Lagrange equations. In particular, if 
$w_\eta=v_\eta\sqrt{\hbar/m\omega}$, $\omega=\omega_{x,y}$, $\omega_z=\lambda\omega_x$, 
$v_{x,y}=v$, $P=\sqrt{2/\pi}Na/\sqrt{\hbar/m\omega}$, and $\tau=\omega t$,
the Euler-Lagrange equations take the following form:
\begin{eqnarray}
&&{d^2\over d\tau^2}v_\eta+\lambda^2_\eta v_\eta={1\over
v^3_\eta}- \nonumber \\
&&P{\partial\over\partial v_\eta}\left[{1\over
v_xv_yv_z}\left(1+\frac{u_2}{g}\int d{\rr} 
\exp\left(-\sum_\eta{\eta^2\over
2v^2_\eta}\right)
{Y_{20}(\theta)\over r^3} \right) \right],
\label{width}
\end{eqnarray}
where $\lambda_{x,y}=1$, $\lambda_z=\lambda$. This equation
describes the motion of a particle with coordinates
$(v_x,v_y,v_z)$ in an effective potential
\begin{eqnarray}
U(v_x,v_y,v_z)&=&{1\over
2}\left(\lambda^2_xv^2_x+\lambda^2_yv^2_y+\lambda^2_zv^2_z\right)\nonumber\\
&+&{1\over
2}\left({1\over v^2_x}+{1\over v^2_y}+{1\over
v^2_z}\right)+{P\over v_xv_yv_z}\nonumber\\
&\times&\left[1+\frac{u_2}{g}\int d{\rr} \exp\left(
-\sum_\eta{\eta^2\over 2v^2_\eta}\right)
{Y_{20}(\theta)\over r^3} \right] \; .
\label{eqv}
\end{eqnarray}
 
\begin{figure}[ht]
\begin{center}\
\epsfxsize=7.0cm
\hspace{0mm}
\psfig{file=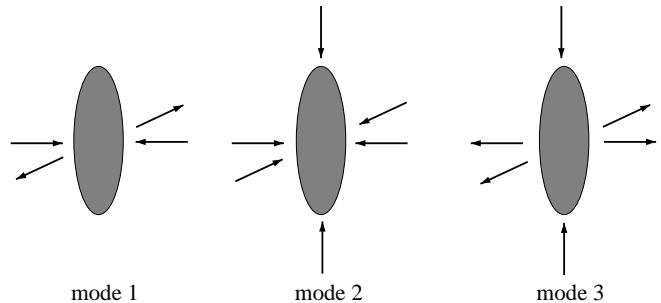,width=\columnwidth,
angle=0,clip}\\[0.1cm]
\end{center}
\caption{Graphical representation of oscillations modes of the
condensate.}
\label{modes}
\end{figure}
Small amplitude
oscillations around the stationary solution 
can be found by studying second derivatives of $U$. This procedure provides 
the excitation frequency for the first three compressional 
excitation modes of the system \cite{Yi2}. The geometry of these modes is depicted in 
Fig.\ \ref{modes}. For cylindrical traps with the axis along the dipole direction, 
the projection of the angular momentum, $m$, on the $z$ axis is a good quantum 
number. For modes $2$ and $3$ we have $m=0$, whereas $m=1$ for mode $1$. 
In the following we call mode $2$ ($3$) as the breathing (quadrupole) 
mode.

\subsection{The nature of the instability} 
\label{sec:exc1b}

To study the nature of the instability one needs to determine which mode 
becomes unstable when the number of atoms reaches the critical value and 
to describe the behavior of the frequency of the 
corresponding mode close to the criticality.
These two issues have been first discussed for the case of short-range interacting 
Bose gases with $a<0$. Bergeman 
\cite{tom} observed numerically that, as the ratio $\gamma$ of the 
nonlinear interaction energy to the trap frequency approaches
a given critical value $\gamma_c$, the frequency of the breathing mode tends to zero and
merges with the frequency of the Goldstone mode corresponding to the overall
phase of the condensate. Above the criticality, the breathing mode 
becomes unstable and attains complex frequency. Singh and Rokhsar \cite{exc_th} 
analyzed this instability using self-similar solutions describing
the modes (equivalently one may employ the variational approach of the 
previous subsection).
They have shown that close to the criticality the frequency of the 
breathing mode $2$ vanishes as $|\gamma -\gamma_c|^{1/4}$.

In the case of a dipolar gas with dominant dipole interactions the situation
is completely different and much more complex. Only for aspect ratios far
above the criticality, $l\gg l^*$ ($l>1.29$) the situation resembles that of 
a gas with $a<0$. The mode corresponding to the lowest
frequency is the breathing mode. This mode becomes unstable when the
parameter $\sigma\to \sigma_c$. The scaling behavior of the frequency of this mode  
can be analyzed employing the variational approach of the previous
section \cite{Yi2}. We find that $\omega_2$ goes to zero as $(\sigma_c-\sigma)^\beta$, 
with $\beta\simeq 1/4$.
For intermediate values of $l>l*$ ($0.75<l<1.29$) the exponent $\beta$ is still $1/4$, but the 
geometry of the mode which drives the instability depends on $\sigma$.
For $\sigma$ far below $\sigma_c$ the mode corresponding to the lowest 
frequency has a breathing 
symmetry, whereas as one approaches the critical value of $\sigma$ the 
lowest mode becomes quadrupole-like.
For $l$ close to $l^*$ ($l<0.75$) the situation changes and the mode corresponding to
the lowest frequency is quadrupolar-like. 
For $l$ not too close to $l^*$ the exponent $\beta$ is still
close to $1/4$. However, as $l$ approaches $l^*$, the exponent $\beta$ departs from $1/4$ towards a
greater value, which becomes $2$ at $l=l_*$. 
The latter prediction must be understood 
as a qualitative argument, since for $l$ very close to $l_*$, the system 
enters a Thomas-Fermi regime where the Gaussian ansatz is no more valid.

\subsection{Numerical results}
\label{sec:exc1c}

In this subsection we study the low-lying collective excitations of a trapped dipolar BEC
by analyzing numerically the response of a BEC after applying an external perturbation.
We compare our numerical results with those obtained from the variational approach described above.
In the following, we employ the method introduced in Ref.\ \cite{Ruprecht} 
for the case of a short-range interacting BEC. After generating the BEC 
ground-state wave function 
using imaginary time evolution of the GPE, we first slightly perturb the 
trapping potential in a periodic way:
\begin{equation}
V({\rr},t)=\frac{1}{2}m\sum_{\eta=x,y,z}
[1+A_{\eta}\sin(\omega_{mod}t+\alpha_{\eta})]^{2} \omega_{\eta}^{2}\eta^2 
\; ,
\end{equation}
where $\omega_{mod}$ is the modulation frequency, 
$A_{\eta}$ are the amplitudes, and $\alpha_{\eta}$ are the initial phases. 
A perturbation of a chosen symmetry can be accomplished 
by properly selecting these parameters. In particular, sufficiently small  
amplitudes are necessary near the instabilities, since the system can be easily 
driven into collapse. On the other hand, by setting large amplitudes 
one may probe various nonlinear effects \cite{Ruprecht,Brewczyk}. 
The response of the system is enhanced by selecting $\omega_{mod}$ in the vicinity of expected
excitation frequencies.

In a second stage, the condensate evolves in an unperturbed trap (i.e. $A_{\eta}=0$). 
The condensate widths are monitored and subsequently Fourier-transformed to reveal the excitation 
spectrum. 
Our aim is to determine the (potentially small) deviations of the excitation frequencies 
with respect to those expected for purely short-range interacting gases, 
as precisely as they can be experimentally measured (typically $\Delta \omega/ \omega_{trap} \simeq 
0.01$ \cite{JILAexc,MITexc}). Long integration times are necessary to accomplish the desired 
spectral resolution. This introduces serious technical difficulties, since 
a large number of integration steps is needed to guarantee the energy 
conservation during the whole evolution. Note that additional 
complications arise from the evaluation of the interaction integral in 
Eq.\ (\ref{GPE1}) at each time step.

\subsubsection{Dominant dipole-dipole interactions}

For the case of dominant dipole-dipole interactions 
\cite{Santos}, the short-range part of Eq.\ (\ref{GPE1}) can be safely omitted. 
We consider two different trap geometries, that were employed at JILA in 
Ref.\ 
\cite{JILAexc}, with a trap aspect ratio $l=8^{-1/4}$ 
(referred to as  pancake JILA trap), and a trap with $l=8^{1/4}$ 
(referred to as  cigar JILA trap). The later trap has been recently 
employed for experiments in $^{85}$Rb \cite{85Rb}.

\begin{figure}[ht]
\begin{center}\
\epsfxsize=6.5cm
\hspace{0mm}
\psfig{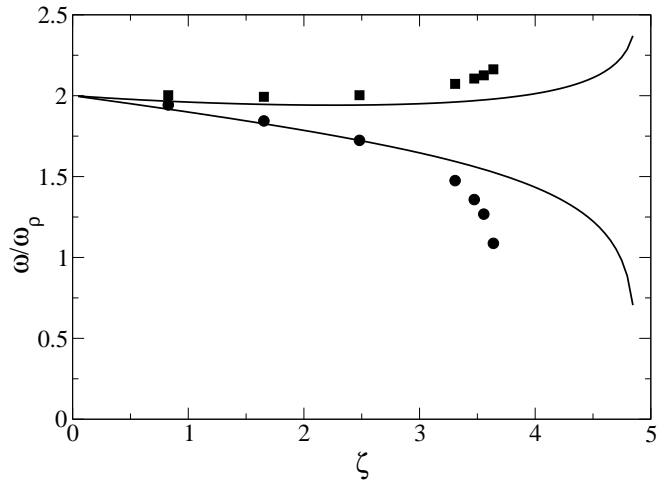}\\[0.1cm]
\end{center}
\caption{Numerical results for the excitation frequencies of modes 1
(squares) and 3 (circles) as functions of the dipolar parameter $\zeta$
for a gas of condensed dipoles in the pancake JILA trap. The solid lines
indicate the corresponding variational results.}
\label{JILApancake}
\end{figure}

In the following, we study the excitation frequencies as a function of the 
dipolar parameter
\begin{equation}
\zeta=N\frac{m}{\hbar^2}\sqrt{\frac{m\omega_{\rho}}{\hbar}}d^2.
\label{zeta_def}
\end{equation}
Note the relation $\sigma=2\sqrt{2\omega_{\rho}/\omega_{min}}\zeta$. Since 
in the pancake JILA trap the breathing mode has a relatively high 
frequency and it is not excited by the perturbation of the GPE, we 
concentrate in this case on the modes $1$ and $3$ of Fig.\ \ref{modes}. 
The results for the pancake JILA trap are presented in Fig. 
\ref{JILApancake}. For 
$\zeta=0$ the ideal-gas result 
$\omega/\omega_{\rho}=2$ is retrieved for both modes $1$ and $3$. 
As $\zeta$ increases mode $1$ is shifted upwards, whereas the frequency 
of mode $3$ (quadrupole) goes down, 
vanishing for a critical $\zeta_{cr}$ at which the system becomes 
unstable. 
Probing the system in this regime is very difficult, since even a slight 
disturbance drives the collapse. For $\zeta>\zeta_{cr}\approx 3.68$ 
it is not possible to obtain stable ground state solutions of the 
time-independent GPE \cite{Santos}. The variational analysis 
reproduces the numerical results for relatively weak 
dipolar interactions. However, it does not describe well the numerical results 
close to the instability. 

\begin{figure}[ht]
\begin{center}\
\epsfxsize=6.5cm
\hspace{0mm}
\psfig{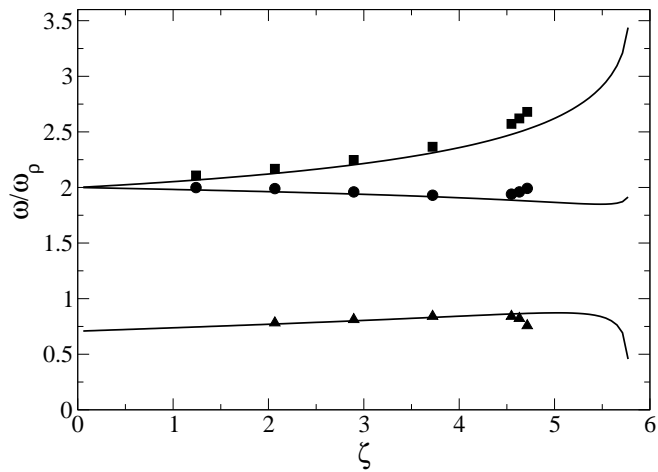}\\[0.1cm]
\end{center}
\caption{Numerical results for the excitation frequencies of modes 1 (squares), 2 (triangles) and 3
(circles) as functions of the dipolar parameter $\zeta$ for a gas of
condensed dipoles in the cigar JILA trap. 
The solid lines indicate the corresponding variational results.}
\label{fig2}
\end{figure}

Fig.~\ref{fig2} depicts a similar dependence for the cigar JILA trap. 
In this case the breathing mode is the lowest one. 
The  frequency of mode $1$ displays an upward shift whereas that of the quadrupole mode
stays essentially untouched. Again, the instability threshold is 
not correctly predicted by the variational analysis. 

\subsubsection{Interplay between short-range and dipole-dipole interactions}

Let us now consider the case in which 
the short-range interactions and the dipole-dipole ones have a  
comparable strength \cite{Yi1,Goral,Martikainen,recyou,Yi2}. 
In particular, we have performed our simulations for the particular case 
of $^{52}$Cr, which has drawn some experimental interest 
\cite{Weinstein,Pfau1,Celotta,Pfau2,EvapCr}, since it possesses a large 
magnetic 
dipole moment of $6 \mu_B$ (Bohr magnetons). The value of $a$ for this 
element is at the moment unknown, and consequently we have explored 
different values of $a$ in the calculations below. 
In the following, the trap parameters 
correspond to those of an ongoing experiment in 
Stuttgart \cite{Pfaupriv}, namely $\omega_z=2 \pi\times 40$ Hz and $\omega_{\rho}=2 \pi\times 485$ Hz. 

\begin{figure}[ht]
\begin{center}\
\epsfxsize=6.5cm
\hspace{0mm}
\psfig{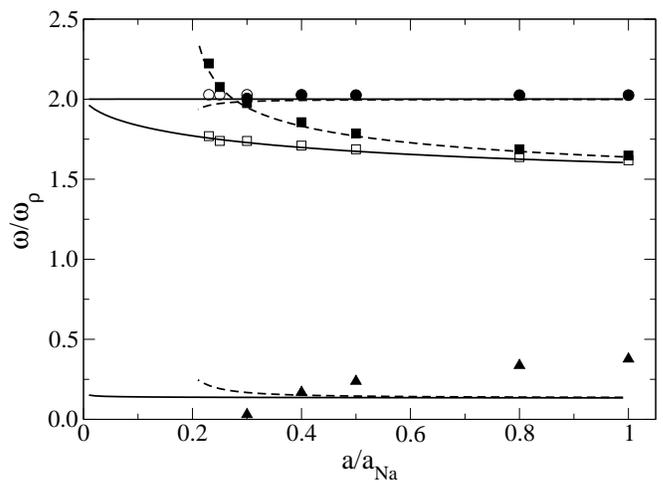}\\[0.1cm]
\end{center}
\caption{Excitation frequencies of modes 1 (filled squares), highest
(filled
triangles) and lowest (filled circles) as functions of the scattering 
length
(expressed in units of the corresponding value for sodium) for a
condensate of $10000$ chromium atoms in a trap with $\omega_z=2 \pi 40$ Hz
and $\omega_{\rho}=2 \pi 485$ Hz - numerical results. The corresponding
variational values are depicted by dashed lines. Solid lines indicate
variational data for the case with no dipole-dipole interactions and
the corresponding numerical results are plotted with empty symbols
(note: we have not been able to probe the lowest mode numerically).}
\label{fig3}
\end{figure}

Fig.~\ref{fig3} shows the three lowest modes for different values of $a$, 
for the case of  $N=10000$ atoms. When $a$ is smaller than a critical $a_{crit}$ 
the system becomes unstable due to an 
unbalanced attractive component of dipole-dipole interactions 
\cite{Yi1,Goral,Martikainen,Yi2}. 
When approaching $a^{crit}/a_{\rm Na}\simeq0.23$ from 
above, the frequency of the lowest mode decreases to $0$ and the system collapses. 
For $0.23<a/a_{Na}<0.26a_{Na}$ the lowest mode (triangles) has a breathing 
geometry, whereas the highest one (circles) is quadrupole-like. The 
opposite is true for $a/a_{Na}>0.26a_{Na}$. The highest mode shows 
virtually no dependence on the scattering length whereas the frequency of 
mode $1$ is shifted upwards. Again, close to the instability, the results 
obtained from the variational calculation differ from those obtained 
numerically. Moreover, the behavior of the lowest mode is 
not reproduced correctly even qualitatively. Fig.~\ref{fig3} also 
presents the excitation frequencies for the same system in the absence of 
dipole-dipole interactions. Apart from the obvious presence of the 
instability in the dipolar case (and the corresponding behavior of the lowest mode) we observe large 
frequency differences for mode $1$ between the cases with and without dipole-dipole interactions. 

The above result suggests that the effects of dipole-dipole interactions 
may be detectable. Having discussed the role of $a$, it is thus important
to discuss possible values of $N$ and $\omega_\rho/\omega_z$  which could 
maximize the dipole-induced frequency shifts. 

\begin{figure}[ht]
\begin{center}\
\epsfxsize=7.0cm
\hspace{0mm}
\psfig{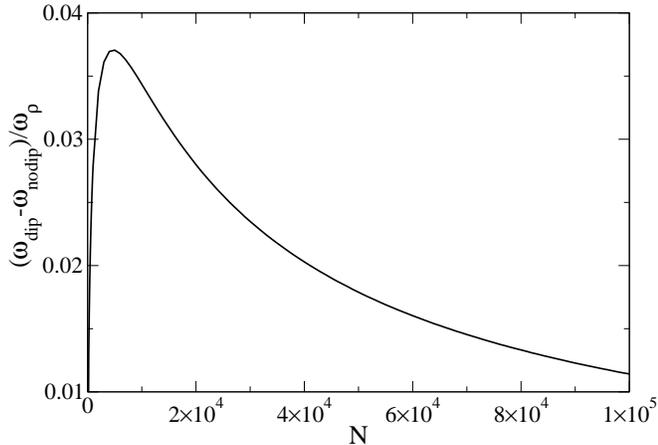}\\[0.1cm]
\end{center}
\caption{Difference of excitation frequency for mode $1$ in the absence
and in the presence of dipole-dipole interactions, as a function of the
total number of atoms $N$, for $a=a_{\rm Na}$ and a trap aspect ratio
$l=\sqrt{485/40}$
(data from the variational analysis).}
\label{fig_vsN}
\end{figure}

Fig.~\ref{fig_vsN} shows the
deviation between the case with and without dipole-dipole interactions,
for the mode $1$, $\omega_z/\omega_{\rho}=40/485$, and $a=a_{Na}$.
For this calculation we have employed the variational method as it is very 
exact away from the instability. We observe that for number of atoms $N$
greater than $5000$ the frequency shift is observable (i.e. greater than
$0.01$), and reaches its maximum ($\gtrsim 3.5\%$) 
for $N\simeq10000$ which should be the number available
in ongoing experiments \cite{Pfaupriv}. In this case, shifts for modes
$2$ and $3$ are substantially lower ($\simeq 0.001$).

\begin{figure}[ht]
\begin{center}\
\epsfxsize=6.5cm
\hspace{0mm}
\psfig{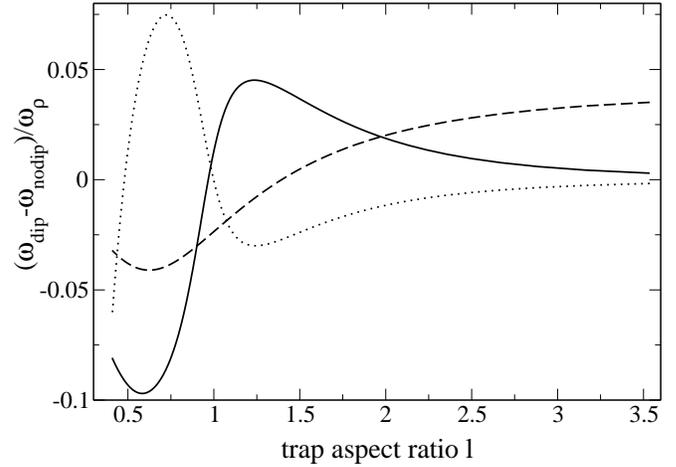}\\[0.1cm]
\end{center}
\caption{Difference of excitation frequency for modes 1 (dashed line), breathing
(dotted line) and quadrupole-like (solid line) between the cases of purely contact and 
mixed (contact and dipole-dipole) interactions as a function of the trap 
aspect ratio for 10000 atoms and $a=a_{\rm Na}$ (data from variational 
analysis).}
\label{vs_asp}
\end{figure}

Finally,  we have analyzed (using the variational method) the dependence 
of the spectrum on the trap aspect ratio $l$ for $a=a_{Na}$ and $N=10000$. 
The results are presented in Fig. \ref{vs_asp}. We observe that in pancake 
traps with $l<0.7$, the quadrupole mode experiences a 
shift of up to $10\%$ with respect to the nondipolar case. We have 
confirmed this conclusion by performing an exact numerical simulation of 
the perturbed GPE for $l=0.58$, $a=a_{Na}$, and 
$N=10000$. For these parameters the excitation frequency of mode $3$ is 
$1.76 \; \omega_{\rho}$ ($1.86\; \omega_{\rho}$) with (without) 
dipole-dipole interactions. The remaining two modes also display large 
shifts over a wide range of aspect ratios.

\section{Ballistic expansion of a dipolar BEC}
\label{sec:exp}

In typical BEC experiments, the measurements are performed after 
removing the trapping potential and allowing for 
a ballistic expansion of the condensate. It is the aim of this section to 
briefly discuss such an expansion, by means of the previously introduced 
variational approach. After evaluating the minimum of the potential 
provided by Eq.\ (\ref{eqv}), i.e. the ground state of the trapped gas, 
we set $\lambda_{x,y,z}=0$ in Eqs.\ (\ref{width}) and evolve these 
equations using the ground state as an initial condition. 
In this section we restrict ourselves to the case of dominant dipole-dipole 
interactions.
 
\begin{figure}[ht]
\begin{center}\
\epsfxsize=5.0cm
\hspace{0mm}
\psfig{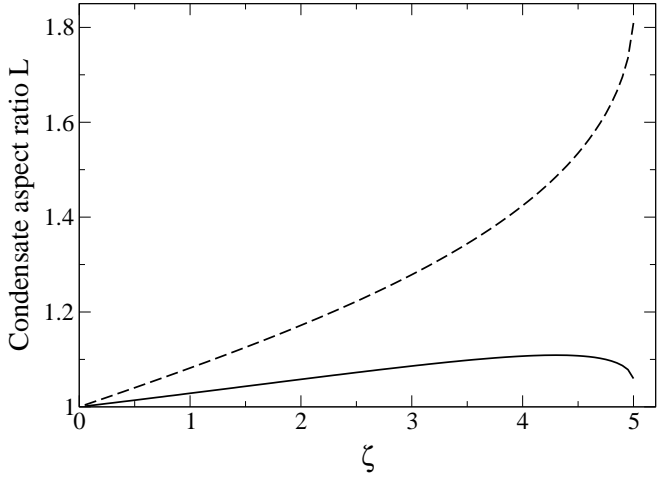}\\[0.1cm]
\end{center}
\caption{Aspect ratio as a function of $\sigma$ for a condensate initially
trapped in a spherical trap of frequency $\omega$, before the expansion
(dashed) and after $48.5\omega^{-1}$ (solid).}
\label{expan}
\end{figure}

The ballistic expansion of a short-range interacting BEC
is characterized by an inversion of the aspect ratio of the condensate 
cloud, i.e. cigar-shaped condensates become pancake-like after expansion, 
and vice versa. This is not necessarily the case in dipolar condensates.
Fig.\ \ref{expan} shows the condensate aspect ratio for a spherical trap, 
before releasing the trap (dashed), and once the aspect ratio reaches 
a stationary value after the expansion (solid). 
The aspect ratio of the cloud decreases during the expansion, 
but never becomes smaller than $1$, i.e. the condensate keeps its 
original cigar-shaped character during the expansion. A similar behavior 
is observed for $1<l<1.12$, where there exits a range of $\zeta$ 
values for which the BEC remains cigar-like. 
For $l_*<l<1$, for $\zeta$ close to $\zeta_{cr}$ the BEC also 
remains cigar-shaped during the time of flight. 
On the contrary, for $l>1.12$ the cloud becomes 
pancake-like after expanding for all $\zeta$, and  
for $l<l_*$ the initial pancake cloud always becomes cigar-shaped.

\section{Ground state of a binary dipolar BEC}
\label{sec:gr2}

In the previous sections we have shown that various novel phenomena are expected 
in dipolar BEC. One can therefore expect that even a richer behavior 
can be displayed by a multicomponent dipolar BEC. 
In the case of two-component dipolar gases, the system properties depend on a 
large number of control parameters, including the 
number of particles in each component, the strengths and 
orientations of the dipoles, and the trap geometries. 
In the following, for simplicity,  we focus on the case of 
a BEC of two dipolar identical components polarized in opposite directions. 
As discussed in Sec.\ \ref{sec:introd} this should be the case for Hund A 
molecules 
in a magnetic field, where the magnetic moment is oriented by the field, 
but the 
(possibly large) electric moment can be parallel or antiparallel to the 
direction of the applied magnetic field. To simplify the analysis even further, 
we consider only a spherical trap. 
Thus, the system is determined by $\zeta$ and $\eta=N_1/N$,
where as before $N$ is the total number of particles and $N_1$ is the 
number of particles in 
the component $1$. The system can be described by a system of two coupled GPEs:

\begin{mathletters}
\begin{eqnarray}
&& i\dot\psi_1= \nonumber \\
&& \left \{ -\frac{\nabla^2}{2}+\frac{r^2}{2}+
+\zeta(\eta V_1({\rr})-(1-\eta)V_2({\rr}))
\right \} \psi_1,\\
&& i\dot\psi_2= \nonumber \\
&&\left \{ -\frac{\nabla^2}{2}+\frac{r^2}{2}+ 
+\zeta(-\eta V_1({\rr})+(1-\eta)V_2({\rr}))
\right \} \psi_2,
\end{eqnarray}
\label{2GPE}
\end{mathletters}
\noindent where $V_i({\rr})=\int d{\rrp} (1-3\cos^2\theta) |\psi_i({\rrp},t)|^2 / |{\rr}-{\rrp}|^3$. 
In the above expression we have employed harmonic oscillator units of 
length $d=\sqrt{\hbar/m\omega}$.
For equal population of the components  ($\eta=0.5$) both ground-state
wave functions are the same. As a  consequence, the nonlocal interaction
terms vanish and the system behaves like an ideal gas,  
the  ground-state wave functions being that of the corresponding harmonic 
oscillator. However, as the interaction strength $\zeta$ increases, 
the system becomes unstable. The reason of this instability is 
rooted in the excitation spectrum and is discussed below.

\begin{figure}[ht]
\begin{center}\
\epsfxsize=7.0cm
\hspace{0mm}
\psfig{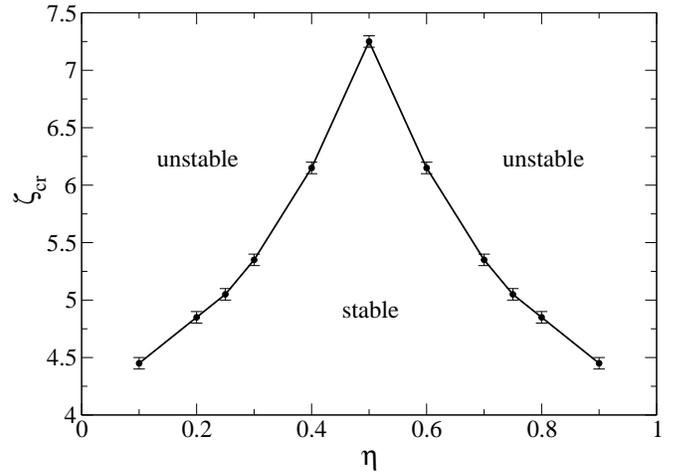}\\[0.1cm]
\end{center}
\caption{Stability diagram for a binary condensate of oppositely polarized
dipolar gases in a spherically symmetric trap. Above the line there are no
stable ground-state solutions.}
\label{fig6}
\end{figure}

For each value of $\eta$ it is possible to compute a critical $\zeta_{cr}$ 
above which no stable solution of Eq. (\ref{2GPE}) exists. The
corresponding stability diagram, resulting from a numerical
solution of Eqs. (\ref{2GPE}), is presented in Fig.~\ref{fig6}.
We observe that the more symmetric the mixture is the more stable it is.

\begin{figure}[ht]
\begin{center}\
\epsfxsize=7.0cm
\hspace{0mm}
\psfig{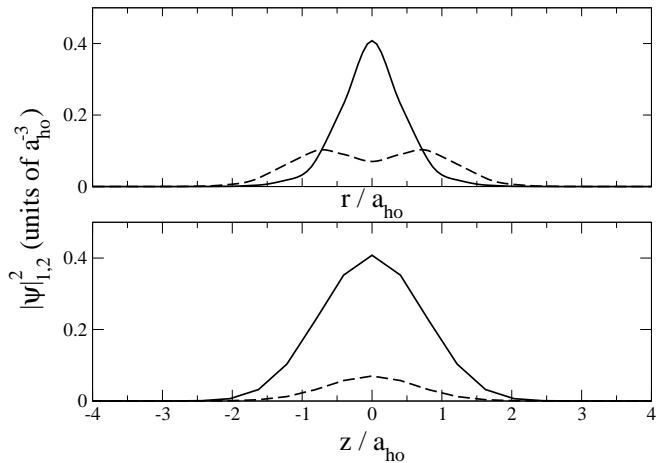}\\[0.1cm]
\end{center}
\caption{Density profiles along the radial (upper graph) and axial
(lower graph) directions for a binary condensate of two oppositely
polarized dipolar gases. Parameters are $\eta=0.8$ and $\zeta=4.8$.}
\label{fig7}
\end{figure}

For $\eta\neq 0.5$ the ground state has a more complicated structure. 
In this respect it resembles solutions for short-range interacting binary condensates
\cite{binary-ground,binaryTrippenbach}.
In particular, the component whose self interaction is stronger partially 
wraps around the other. 
However, contrary to short-range interacting binary BEC, due to the anisotropy of the 
dipole-dipole interactions, the density dip 
in one component appears only along the radial direction, whereas  
along $z$ both components possess Gaussian profiles. 
An example of such a structure is shown in Fig.~\ref{fig7}.

In the analysis presented in this section we have neglected the influence 
of short-range interactions. Since the two components are identical except 
for the opposite dipole moments, the short-range interactions (both 
intercomponent and the intracomponent ones) are the same. Therefore, in 
the absence of dipole-dipole interactions a miscible mixture is expected, 
which differs from the ideal-gas Gaussian case considered above 
(eventually acquiring a Thomas-Fermi profile). However, in the same way as 
above, the dipole-dipole mean fields will be exactly canceled for 
$\eta=0.5$. Therefore, the cloud is expected to remain in an unperturbed 
Thomas-Fermi profile until the $\zeta$ coefficient reaches a critical value 
$\zeta_{cr}$ (only quantitatively different from that obtained in 
absence of short-range interactions). For $\zeta>\zeta_{cr}$ the system 
becomes unstable.

\section{Excitations of a binary dipolar BEC}
\label{sec:exc2}

In this section we analyze the excitation spectrum of the previously 
discussed gas of two oppositely oriented dipolar components. 
We first discuss the homogeneous case, and then study 
the case of harmonically trapped binary dipolar BEC. 

\subsection{Homogeneous space}

Let us consider two species of dipoles ${\bf d}_1=d_1{\bf u}_z$ 
and ${\bf d}_2=-d_2{\bf u}_z$ with respective densities $n_1$ and $n_2$
in 
a homogeneous box of volume $V$. The Hamiltonian in second quantization is:
\begin{eqnarray}
&&\hat H=\sum_{\kk\alpha=1,2}\frac{\hbar^2\kk^2}{2m}a_{\kk\alpha}^{\dag}a_{\kk\alpha}
\nonumber \\
&&+\frac{1}{2}\sum_{\alpha=1,2}\sum_{\kk_1,\kk_2,\kk_1',\kk_2'}
g(\kk_1',\kk_2';\kk_1,\kk_2)
a_{\kk_1'\alpha}^{\dag}a_{\kk_2'\alpha}^{\dag}a_{\kk_2\alpha}a_{\kk_1\alpha} 
\nonumber \\
&&-\sum_{\kk_1,\kk_2,\kk_1',\kk_2'}
g(\kk_1',\kk_2';\kk_1,\kk_2)
a_{\kk_1'1}^{\dag}a_{\kk_2'2}^{\dag}a_{\kk_22}a_{\kk_11},
\end{eqnarray}
with 
\begin{eqnarray}
&&g(\kk_1',\kk_2';\kk_1,\kk_2)=\nonumber \\
&&\int d\rr \int d\rrp \psi_{\kk_1'}^*(\rr)
\psi_{\kk_2'}^*(\rrp)V(\rr-\rrp)\psi_{\kk_2}(\rrp)\psi_{\kk_1}(\rr),
\end{eqnarray}
and $a_{\kk i}$ the annihilation operator of a particle of the 
$i$-th component with momentum $\kk$. Performing the Bogoliubov approximation 
($a_{01}^{\dag}=a_{01}=\sqrt{N_1}$ and 
$a_{02}^{\dag}=a_{02}=\sqrt{N_2}$), one obtains 
\begin{eqnarray}
&&\hat H=\sum_{\kk\not= 0}\{
[\epsilon_\kk+2B(\hat \kk)]a_{\kk 1}^{\dag}a_{\kk 1}
+[\epsilon_\kk+2\lambda^2B(\hat \kk)]a_{\kk 2}^{\dag}a_{\kk 2}
\nonumber \\
&&+B(\hat \kk)
[(a_{\kk 1}^{\dag}a_{-\kk 1}^{\dag}+a_{\kk 1}a_{-\kk 1})
+\lambda^2(a_{\kk 2}^{\dag}a_{-\kk 2}^{\dag}+a_{\kk 2}a_{-\kk 2})
\nonumber \\
&&-2\lambda(a_{\kk 1}a_{-\kk 2}+a_{\kk 1}^{\dag}a_{-\kk 2}
+a_{\kk 1}^{\dag}a_{\kk 2}+a_{\kk 2}^{\dag}a_{\kk 1})]\},
\end{eqnarray}
with $\epsilon_\kk=\hbar^2\kk^2/2m$,
$\sqrt{n_2d_2^2}=\lambda\sqrt{n_1d_1^2}$ and 
$B(\hat \kk)=\frac{8\pi}{3}\sqrt{\frac{\pi}{5}}
(n_1d_1^2)Y_{20}(\hat \kk)$.
The diagonalization of the Hamiltonian provides  
two different branches of quasiparticle excitations:
\begin{mathletters}
\begin{eqnarray}
&&E_1=\epsilon_\kk, \label{soft} \\
&&E_2=\sqrt{\epsilon_\kk^2+\frac{32\pi}{3}\sqrt{\frac{\pi}{5}}
(1+\lambda^2)(n_1d_1^2)Y_{20}(\hat \kk)\epsilon_\kk} \label{unst}.
\end{eqnarray}
\end{mathletters}
For the case of $\lambda\rightarrow 0$ the 
spectrum of a homogeneous single-component dipolar BEC is 
discussed in Sec.\ \ref{sec:gr1}. 
As observed, the spectra (\ref{soft}) is that of a soft mode, whereas 
the second branch (\ref{unst}) is even more unstable than that of a single 
component BEC.

\subsection{Trapped case}

For simplicity we limit ourselves to the case of a spherical trap 
with frequency $\omega$, but similar arguments can 
be employed for a cylindrical trap. We also consider 
$|{\bf d}_1|=|{\bf d}_2|$ and $\eta=0.5$ (i.e. $N_1=N_2=N/2$). 
As we have already noted, in this case the ground state of the system is
the ground state of a harmonic oscillator. It is therefore more convenient 
to rewrite the Hamiltonian in terms of the creation and annihilation 
operators of the harmonic oscillator modes for the dipole components $1$
and $2$:
\begin{eqnarray}
&& \hat H=\sum_\nn\hbar\omega_\nn(a_{\nn1}^{\dag}a_{\nn1}+a_{\nn2}^{\dag}a_{\nn2})
+ \frac{N_1d^2}{2}\sum_{\nn_1,\nn_2\not= 0} \nonumber \\
&&
\{ g(\nn_1,\nn_2;0,0)a_{\nn_11}^{\dag}a_{\nn_21}^{\dag}
+g(0,0;\nn_1,\nn_2)a_{\nn_11}a_{\nn_21}
\nonumber \\
&& +g(\nn_1,\nn_2;0,0)a_{\nn_12}^{\dag}a_{\nn_22}^{\dag}
+g(0,0;\nn_1,\nn_2)a_{\nn_12}a_{\nn_22}
\nonumber \\
&&+2g(\nn_1,0;\nn_2,0)a_{\nn_11}^{\dag}a_{\nn_21}
+2g(\nn_1,0;\nn_2,0)a_{\nn_12}^{\dag}a_{\nn_22}
\nonumber \\
&&-2g(0,0;\nn_1,\nn_2)a_{\nn_11}a_{\nn_22}
-2g(\nn_1,\nn_2;0,0)a_{\nn_11}^{\dag}a_{\nn_22}^{\dag}
\nonumber \\
&&-2g(\nn_1,0;0,\nn_2)a_{\nn_11}^{\dag}a_{\nn_22}
-2g(\nn_1,0;0,\nn_2)a_{\nn_12}^{\dag}a_{\nn_21} \},
\end{eqnarray}
where $\nn\equiv (n,l,m)$ 
denotes the set of spherical quantum numbers 
for the corresponding eigenstate of the harmonic oscillator, 
$\hbar\omega_\nn$ is the energy of the eigenstate $\nn$, and 
\begin{eqnarray}
&&g(\nn_1,\nn_2;\nn_3,\nn_4)=\nonumber \\
&&\int dr \int dr' \psi_{\nn_1}^*(r)
\psi_{\nn_2}^*(r')V(r-r')\psi_{\nn_3}(r')\psi_{\nn_4}(r).
\end{eqnarray}
Let us perform the Bogoliubov transformation:
\begin{equation}
b_\nu=\sum_{\nn\not= 0}(u_{\nu\nn}a_{\nn 1}+v_{\nn \nu}a_{\nn 1}^{\dag}+
\tilde u_{\nu\nn}a_{\nn 2}+\tilde v_{\nn \nu} a_{\nn 2}^{\dag} ),
\end{equation}
which satisfies $\epsilon_\nu b_\nu=[b_\nu,\hat H]$, leading to the 
corresponding Bogoliubov-de Gennes (BdG) equations. 
Such equations can be simplified 
by taking the more appropriate variables 
$A_{\nu\nn}=u_{\nu\nn}+\tilde u_{\nu\nn}$,
$B_{\nu\nn}=v_{\nu\nn}+\tilde v_{\nu\nn}$,
$C_{\nu\nn}=u_{\nu\nn}-\tilde u_{\nu\nn}$,
$D_{\nu\nn}=v_{\nu\nn}-\tilde v_{\nu\nn}$. Then, the BdG equations take the form:
\begin{mathletters}
\begin{eqnarray}
&&\epsilon_\nu A_{\nu\nn}=\hbar\omega_\nn A_{\nu\nn}, \\
&&\epsilon_\nu B_{\nu\nn}=-\hbar\omega_\nn B_{\nu\nn}, \\
&&\epsilon_\nu C_{\nu\nn}=\hbar\omega_\nn C_{\nu\nn} \nonumber \\
&&+2N_1d^2
\sum_{\nnp\not=0}\{g(\nnp,0;\nn,0)C_{\nu\nnp}-g(0,0;\nn,\nnp)D_{\nu\nnp}\}, \\
&&\epsilon_\nu D_{\nu\nn}=-\hbar\omega_\nn D_{\nu\nn} \nonumber \\
&&+2N_1d^2
\sum_{\nnp\not=0}\{g(\nn,\nnp;0,0)C_{\nu\nn}-g(\nn,0;\nnp,0)D_{\nu\nnp}\}.
\end{eqnarray}
\end{mathletters}
In the following, we omit for simplicity $\nu$ in the notation for the $A$, 
$B$, $C$ and $D$ coefficients.
We observe that, like in the homogeneous case, we have two different 
branches of quasi-particles, one of them being a soft mode. 
The last two equations are the BdG equations for the non-soft mode. 
From the properties of spherical harmonics and the $3j$-symbols,  
it is possible to obtain that 
$g(0,0;nlm,n'l'm')=g(nlm,n'l'm';0,0)=
g(0,0;nlm,n'l'-m)\delta_{m',-m}$
$g(nlm,0;n'l'm',0)=g(n'l'm',0;nlm,0)=
g(0,0;nlm,n'l'-m)\delta_{m',m}$. 
Also, $l'=l,l\pm 2$ must be fulfilled.
Employing these properties, and changing to the variables 
$F_{nlm}=C_{nlm}-(-1)^mD_{nl-m}$,
$G_{nlm}=C_{nlm}+(-1)^mD_{nl-m}$, one obtains that 
$G_{nlm}=\frac{\epsilon}{\hbar\omega_n}F_{nlm}$, and 
\begin{eqnarray}
&&\epsilon^2F_{nlm}=(\hbar\omega_n)^2F_{nlm} \nonumber \\
&&+
4N_1d^2\hbar\omega_n\sum_{n\not= 0}g(0,0;nlm,n'l'-m)F_{n'l'm}.
\end{eqnarray}
One can write: 
\begin{eqnarray}
&&g(0,0;nlm,n'l'm')=\nonumber \\
&&(-1)^{(l'-l)/2}\frac{1}{2\sqrt{2}a_0^3}
h(n,l;n',l')\langle l'm|l,m|20\rangle,
\label{g00}
\end{eqnarray}
where $a_0=\sqrt{\hbar/2m\omega}$. In Eq.\ (\ref{g00})
the angular contribution is given by 
\begin{eqnarray}
&&\langle l',m|lm|20\rangle = \nonumber \\
&&(-1)^m 
{\, l'\, l\, 2 \choose -mm0}
{\, l'\, l\, 2 \choose 000}
\left [ \frac{5(2l+1)(2l'+1)}{4\pi} \right ]^{1/2},
\end{eqnarray}
whereas the coefficients $h(n,l;n',l')$ constitute the radial contribution 
and are of the form:
\begin{eqnarray}
&&h(n,l;n',l')=\frac{8}{3\sqrt{5}}\frac{1}{2^{(n+n')/2}} \nonumber \\
&&\times \frac{\Gamma\left(\frac{n+n'+3}{2}\right)}
{\sqrt{\Gamma\left(\frac{n+l+3}{2}\right)
\Gamma\left(\frac{n-l+2}{2}\right)
\Gamma\left(\frac{n'+l'+3}{2}\right)
\Gamma\left(\frac{n'-l'+2}{2}\right)
}}.
\end{eqnarray}

Then the problem reduces to finding the eigenvalues of the following
equation:
\begin{eqnarray}
&&\epsilon^2F_{nlm}=n^2F_{nlm}+\sqrt{2}\zeta n \nonumber \\
&&\sum_{l'= l,l\pm 2}\sum_{n'\ge l'}(-1)^{(l-l')/2}h(nl;n'l')
\langle l'm|lm|20 \rangle F_{n'l'm} \; ,
\end{eqnarray}
where $\zeta$ is defined in Eq. (\ref{zeta_def}), and $\epsilon$ is in units 
of $(\hbar\omega)^2$ . 
From the analysis of the spectrum one can observe that for 
$\zeta>\zeta_{cr}=6.7$ 
the energy of the lowest state of the subspace $n=2$, $m=0$ becomes imaginary. 
Therefore, although the ground-state of the mixture of antiparallel dipolar 
components is that of an ideal gas, the system will eventually 
become unstable for $\eta>\eta_{cr}$ due to the appearance of imaginary 
excitation frequencies. The value of $\zeta_{cr}$ 
compares very well with the numerically 
found value of $\zeta_{cr}\simeq 7.3$ for the onset of the instability.

Finally, it is interesting to analyze the situation in which 
$\zeta\ll 1$, 
since in this situation the states with different $n$ do 
not mix, and one can analytically diagonalize in each $n$ subspace.
The first excited states are (in units of $(\hbar\omega)^2$):
\begin{eqnarray}
&&\epsilon^2 (n=1, l=1, m=0)=1+4\sqrt{2/\pi}\zeta/15, \\
&&\epsilon^2 (n=1, l=1, m=\pm 1)=1-2\sqrt{2/\pi}\zeta/15, \\
&&\epsilon^2 (n=2, l=\{2,0\}, m=0)=4+(2\pm\sqrt{102}) \sqrt{2/\pi}\zeta /21, \\ 
&&\epsilon^2 (n=2, l=2, m=\pm 1)=4+2\sqrt{2/\pi}\zeta/21.
\end{eqnarray}
From these expressions it becomes clear that, for example, in the subspace 
of states with $n=2$, $m=0$ (which is an admixture of the noninteracting 
quadrupole and monopole modes) even for $\zeta\sim 0.1$ deviations of more 
than $1\%$ are expected. One can also observe that vortices with vorticity 
in the dipole direction are less energetic than those with vorticity in 
the plane perpendicular to the dipole direction.

\section{conclusions}
\label{sec:concl}

In this paper we have analyzed the ground state and the excitation
spectrum of single- and two-component dipolar condensates.
For the case of single-component BECs we have  
shown that their stability properties are determined by  the
trapping geometry. In particular, for sufficiently pancake-like traps
the condensate is always stable independent of the number of particles. We
have then analyzed the excitation spectrum of the 
single-component condensate by analyzing
the response of the condensate to small perturbations and comparing 
the results with analytical calculations based on a variational approach. 
We have discussed in detail the nature of the instability 
and associated it with vanishing of frequency of one of the excitation modes.
The scaling behavior of this frequency was also analyzed. 
We have discussed different possible scenarios in which 
the dipole-dipole effects could have observable effects 
even in the presence of a dominant 
short-range interaction. This analysis could be of special interest
for ongoing experiments on atoms with large magnetic moments, such as 
chromium \cite{Pfaupriv}. We have provided guidelines for experimental
parameters corresponding to largest discrepancies between the cases
with and without dipole-dipole interactions.
In the second part of the paper we have analyzed the properties of a
two-component BEC of dipolar particles. In particular, we have studied 
the stability of the ground state as a function of the relative density
of both species as well as the appearance of phase separation 
in the binary condensate. Finally, we have obtained the
excitation spectrum for this particular physical system and discussed 
the nature of its instability.

We acknowledge support from the Alexander Von Humboldt Foundation, the ZIP 
Program of the German Government, the Deutscher Akademischer 
Austauschdienst (DAAD), the Deutsche Forschungsgemeinschaft, the RTN Cold 
Quantum Gases, ESF Program BEC2000+, and the subsidy of the Foundation for 
Polish Science. K.G. acknowledges support by the Polish KBN grant 5 P03B 
102 20. We thank M. Gajda, M. Lewenstein, G. Meijer, T. Pfau, K. 
Rz\c{a}\.{z}ewski, G. V. Shlyapnikov, E. Tiemann and L. You for fruitful 
discussions. Part of the results was obtained using computers at the 
Interdisciplinary Center for Mathematical and Computational Modeling at 
Warsaw University. K. G. and L. S. are grateful for the kind hospitality 
extended to them in Hannover and Warsaw, respectively.


\begin{references}

\bibitem{BEC}
M.H. Anderson, J.R. Ensher, M.R. Matthews, C.E. Wieman, and E.A. Cornell,
Science {\bf 269}, 198 (1995); K.B. Davis, M.-O. Mewes, M.R. Andrews, N.J. 
van Druten, D.S. Durfee, D.M. Kurn, and W. Ketterle, Phys. Rev. Lett. {\bf 
75}, 3969 (1995); D.G. Fried, T.C. Killian, L. Willmann, D. Landhuis, S.C. 
Moss, D. Kleppner, and T.J. Greytak, {\it ibid.} {\bf 81}, 3811 (1998); 
S.L. Cornish, N.R. Claussen, J.L Roberts, E.A. Cornell, and C.E. Wieman,
{\it ibid.} {\bf 85}, 1795 (2000); A. Robert, O. Sirjean, A. Browaeys, J. 
Poupard, S. Nowak, D. Boiron, C.I. Westbrook, and A. Aspect, Science {\bf 
292}, 461 (2001); F. Pereira Dos Santos, J. Leonard, J. Wang, C.J. 
Barrelet, F. Perales, E. Rasel, C.S. Unnikrishnan, M. Leduc, and C. 
Cohen-Tannoudji, Phys. Rev. Lett. {\bf 86}, 3459 (2001); G. Modugno, G. 
Ferrari, G. Roati, R.J. Brecha, A. Simoni, and M. Inguscio, Science {\bf 
294}, 1320 (2001).

\bibitem{Li}
C.C. Bradley, C.A. Sackett, J.J. Tollett, and R.G. Hulet, Phys. Rev. Lett.
{\bf 75}, 1687 (1995) and Erratum {\bf 79}, 1170(E) (1997).

\bibitem{Dalfovo}
see, for example, F. Dalfovo, S. Giorgini, L.P. Pitaevskii, and S. 
Stringari, Rev. Mod. Phys. {\bf 71}, 463 (1999).

\bibitem{collapse}
J.M. Gerton, D. Strekalov, I. Prodan, and  R.G. Hulet, Nature (London) 
{\bf 408}, 692 (2000).

\bibitem{keith} 
P.A. Ruprecht, M.J. Holland, K. Burnett, and M. Edwards, Phys. Rev. A {\bf 
51}, 4704 (1995).

\bibitem{baym} 
G. Baym and C.J. Pethick, Phys. Rev. Lett. {\bf 76}, 6 (1996). 

\bibitem{negat}
Y. Kagan, G.V. Shlyapnikov, and J.T.M. Walraven, Phys. Rev. Lett. {\bf
76}, 2670 (1996).  

\bibitem{Yi1}
S. Yi and L. You, Phys. Rev. A {\bf 61}, 041604 (2000).

\bibitem{Goral}
K. G\'oral, K. Rz\c{a}\.zewski, and T. Pfau, Phys. Rev. A {\bf 61}, 051601
(2000).

\bibitem{Santos}
L. Santos, G.V. Shlyapnikov, P. Zoller, and M. Lewenstein, Phys. Rev. 
Lett. {\bf 85}, 1791 (2000).

\bibitem{Yi2}
S. Yi and L. You, Phys. Rev. A {\bf 63}, 053607 (2001).

\bibitem{recyou} 
S. Yi and  L. You, Phys. Rev. A {\bf 66}, 013607 (2002).

\bibitem{Martikainen}
J.-P. Martikainen, M. Mackie, and K.-A. Suominen, Phys. Rev. A {\bf 64},
037601 (2001).

\bibitem{Meystre}
H. Pu, W. Zhang, and P. Meystre, Phys. Rev. Lett. {\bf 87}, 140405 (2001);
W. Zhang, H. Pu, C. Search, and P. Meystre, {\it ibid.} {\bf 88}, 
060401 (2002).

\bibitem{Giovanazzi}
S. Giovanazzi, D. O'Dell, and G. Kurizki, Phys. Rev. Lett. {\bf 88}, 
130402 (2002);  J. Phys. B {\bf 34}, 4757 (2001). 

\bibitem{kurizki}
D. O'Dell, S. Giovanazzi, G. Kurizki, and V.M. Akulin, Phys. Rev. Lett.
{\bf 84}, 5687 (2000); S. Giovanazzi, D. O'Dell, and G. Kurizki, Phys.
Rev. A {\bf 63}, 031603 (2001); S. Giovanazzi, G. Kurizki, I.E. Mazets,
and S. Stringari, Europhys. Lett. {\bf 56}, 1 (2001); T.K. Ghosh, Phys. 
Rev. A {\bf 65}, 053616 (2002).

\bibitem{molecules}
J.D. Weinstein, R. deCarvalho, T. Guillet, B. Friedrich, and J.M. Doyle, 
Nature (London) {\bf 395}, 148 (1998); A. Fioretti, D. Comparat, A.
Crubellier, O. Dulieu, F. Masnou-Seeuws, and P. Pillet, Phys. Rev. Lett.
{\bf 80}, 4402 (1998); T. Takekoshi, B.M. Patterson, and R.J. Knize, {\it
ibid.} {\bf 81}, 5105 (1998); A.N. Nikolov, E.E. Eyler, X.T. Wang, J. Li, 
H. Wang, W.C. Stwalley, and P.L. Gould, {\it ibid.} {\bf 82}, 703 (1999);
H.L. Bethlem, G. Berden, and G. Meijer, {\it ibid.} {\bf 83}, 1558 (1999);
A.N. Nikolov, J.R. Ensher, E.E. Eyler, H. Wang, W.C. Stwalley, and
P.L Gould, {\it ibid.} {\bf 84}, 246 (2000); C. Gabbanini, A. Fioretti, A.
Lucchesini, S. Gozzini, and M. Mazzoni, {\it ibid.} {\bf 84}, 2814 (2000);
H.L. Bethlem, G. Berden, F.M.H. Crompvoets, R.T. Jongma, A.J.A. van Roij, 
and G. Meijer, Nature (London) {\bf 406}, 491 (2000); F.M.H. Crompvoets, 
H.L. Bethlem, R.T. Jongma, and G. Meijer, {\it ibid.} {\bf 411}, 174 
(2001).

\bibitem{Heinzen} 
R. Wynar, R.S. Freeland, D.J. Han, C. Ryu, and D.J. Heinzen, Science
{\bf 287}, 1016 (2000).

\bibitem{Weinstein}
J.D. Weinstein, R. deCarvalho, J. Kim, D. Patterson, B. Friedrich, and
J.M. Doyle,  Phys. Rev. A {\bf 57}, R3173 (1998).

\bibitem{Pfau1}
A.S. Bell, J. Stuhler, S. Locher, S. Hensler, J. Mlynek, and T. Pfau,
Europhys. Lett. {\bf 45}, 156 (1999).

\bibitem{Celotta}
C.C. Bradley, J.J. McClelland, W.R. Anderson, and R.J. Celotta, Phys. Rev.
A {\bf 61}, 053407 (2000).

\bibitem{Pfau2}
J. Stuhler, P. O. Schmidt, S. Hensler, J. Werner, J. Mlynek, and T. Pfau, 
Phys. Rev. A {\bf 64}, 031405 (2001).

\bibitem{EvapCr}
J.D. Weinstein, R. deCarvalho, C.I. Hancox, and J.M. Doyle, Phys. Rev. A 
{\bf 65}, 021604 (2002).

\bibitem{europium}
J. Kim, B. Friedrich, D.P. Katz, D. Patterson, J.D. Weinstein, R.
DeCarvalho, and J.M. Doyle, Phys. Rev. Lett. {\bf 78}, 3665 (1997).

\bibitem{FeshbachMIT}
S. Inouye, M.R. Andrews, J. Stenger, H.J. Miesner, D.M. Stamper-Kurn, 
and W. Ketterle, Nature (London) {\bf 392}, 151 (1998).

\bibitem{85Rb}
J.L. Roberts, N.R. Claussen, S.L. Cornish, E.A. Donley, E.A. Cornell, and
C.E. Wieman, Phys. Rev. Lett. {\bf 86}, 4211 (2001).

\bibitem{Lukin}
M.D. Lukin, M. Fleischhauer, R. C\^{o}t\'{e}, L.M. Duan, D. Jaksch, J.I. 
Cirac, and P. Zoller, Phys. Rev. Lett. {\bf 87}, 037901 (2001).

\bibitem{JILAexc}
D.S. Jin, J.R. Ensher, M.R. Matthews, C.E. Wieman, and E.A. Cornell, Phys.
Rev. Lett. {\bf 77}, 420 (1996).

\bibitem{MITexc}
M.-O. Mewes, M.R. Andrews, N.J. van Druten, D.M. Kurn, D.S. Durfee, C.G.
Townsend, and W. Ketterle, Phys. Rev. Lett. {\bf 77}, 988 (1996).

\bibitem{exc_th}
K.G. Singh and D.S. Rokhsar, Phys. Rev. Lett. {\bf 77}, 1667 (1996); M. 
Edwards, P.A. Ruprecht, K. Burnett, R.J. Dodd, and C.W. Clark, {\it ibid.} 
{\bf 77}, 1671 (1996); S.Stringari, {\it ibid.} {\bf 77}, 2360 (1996); L. 
You, W. Hoston, and M. Lewenstein, Phys. Rev. A {\bf 55}, R1581 (1997); P. 
Ohberg, E.L. Surkov, I. Tittonen, S. Stenholm, M. Wilkens, and G.V. 
Shlyapnikov, {\it ibid.} {\bf 56}, R3346 (1997).

\bibitem{Perez}   
V.M. P\'{e}rez-Garc\'{\i}a, H. Michinel, J.I. Cirac, M. Lewenstein, and P.
Zoller, Phys. Rev. Lett. {\bf 77}, 5320 (1996).


\bibitem{tom} T. Bergeman, Phys. Rev. A {\bf 55}, 3658 (1997).

\bibitem{binaryJILA}
C.J. Myatt, E.A. Burt, R.W. Ghrist, E.A. Cornell, and C.E. Wieman, Phys.
Rev. Lett. {\bf 78}, 586 (1997); D.S. Hall, M.R. Matthews, J.R. Ensher,
C.E. Wieman, and E. A. Cornell, {\it ibid.} {\bf 81}, 1539 (1998); D.S.
Hall, M.R. Matthews, C.E. Wieman, and E.A. Cornell, {\it ibid.} {\bf
81}, 1543 (1998).

\bibitem{binaryMIT}
D.M. Stamper-Kurn, M.R. Andrews, A.P. Chikkatur, S. Inouye, H.J. Miesner,
J. Stenger, and W. Ketterle, Phys. Rev. Lett. {\bf 80}, 2027 (1998); J.
Stenger, S. Inouye, D.M. Stamper-Kurn, H.J. Miesner, A.P. Chikkatur, and
W. Ketterle, Nature (London) {\bf 396}, 345 (1998); H.J. Miesner, D.M.
Stamper-Kurn, J. Stenger, S. Inouye, A.P. Chikkatur, and W. Ketterle,
Phys. Rev. Lett. {\bf 82}, 2228 (1999).

\bibitem{binary-ground}
T.L. Ho and V.B. Shenoy, Phys. Rev. Lett. {\bf 77}, 3276 (1996); B.D. 
Esry, C.H. Greene, J.P. Burke, and J.L. Bohn, {\it ibid.} {\bf 78},
3594 (1997); P. \"{O}hberg and S. Stenholm, Phys. Rev. A {\bf 57}, 1272
(1998); E. Timmermans, Phys. Rev. Lett. {\bf 81}, 5718 (1998).

\bibitem{binaryTrippenbach}
M. Trippenbach, K. G\'{o}ral, Y. Band, B. Malomed, and K.
Rz\c{a}\.{z}ewski, J. Phys. B {\bf 33}, 4017 (2000). 

\bibitem{binary-exc}
Th. Busch, J.I. Cirac, V.M. P\'{e}rez-Garc\'{\i}a, and P. Zoller,  
Phys. Rev. A {\bf 56}, 2978 (1997); B.D. Esry and C.H. Greene, {\it ibid.} 
{\bf 57}, 1265 (1998); P. \"{O}hberg and S. Stenholm, J. Phys. B {\bf
32}, 1959 (1999).

\bibitem{Meijerpriv} 
G. Meijer, private communication.

\bibitem{Pfaupriv}
T. Pfau, private communication.

\bibitem{soliton} 
V.E. Zakharov, S.V. Manakov, S.P. Novikov, and L.V. Pitaevskii, {\em
Teoria solitonov: Metod obratnoy zadatchi}, (Nauka, Moscow, 1980).

\bibitem{self-similar}
E. Timmermans, R. C\^{o}t\'{e}, and I. Simbotin, J. Phys B {\bf 33}, 4157
(2000).

\bibitem{Ruprecht}
P.A. Ruprecht, M. Edwards, K. Burnett, and C.W. Clark, Phys. Rev. A {\bf 
54}, 4178 (1996).

\bibitem{Brewczyk}
M. Brewczyk, K. Rz\c{a}\.{z}ewski, and C.W. Clark, Phys. Rev. A {\bf
57}, 488 (1998).

\end{references}
\end{document}